\documentclass{aa}
\usepackage{natbib}
\usepackage{graphicx}
\usepackage{txfonts}

\begin{document}

\title{GRB060614: a ``fake'' short GRB from a merging binary system}

\author{
L. Caito\inst{1,2}
\and
M.G. Bernardini\inst{1,2}
\and
C.L. Bianco\inst{1,2}
\and
M.G. Dainotti\inst{1,2}
\and
R. Guida\inst{1,2}
\and
R. Ruffini\inst{1,2,3}
}

\institute{
Dipartimento di Fisica and ICRA, Universit\`a di Roma ``La Sapienza'', Piazzale Aldo Moro 5, I-00185 Roma, Italy.
\and
ICRANet, Piazzale della Repubblica 10, I-65122 Pescara, Italy. E-mails: letizia.caito@icra.it; maria.bernardini@icra.it; bianco@icra.it; dainotti@icra.it; roberto.guida@icra.it; ruffini@icra.it
\and
ICRANet, Universit\'e de Nice Sophia Antipolis, Grand Ch\^ateau, BP 2135, 28, avenue de Valrose, 06103 NICE CEDEX 2, France.
}

\titlerunning{GRB060614: a ``fake'' short GRB from a merging binary system}

\authorrunning{Caito et al.}

\date{}

\abstract{
GRB060614 observations by VLT and by Swift have infringed the traditionally accepted gamma-ray burst (GRB) collapsar scenario that purports the origin of all long duration GRBs from supernovae (SN). GRB060614 is the first nearby long duration GRB clearly not associated with a bright Ib/c SN. Moreover, its duration ($T_{90} \sim 100$ s) makes it hardly classifiable as a short GRB. It presents strong similarities with GRB970228, the prototype of a new class of ``fake'' short GRBs that appear to originate from the coalescence of binary neutron stars or white dwarfs spiraled out into the galactic halo.
}
{
Within the ``canonical'' GRB scenario based on the ``fireshell'' model, we test if GRB060614 can be a ``fake'' or ``disguised'' short GRB. We model the traditionally termed ``prompt emission'' and discriminate the signal originating from the gravitational collapse leading to the GRB from the process occurring in the circumburst medium (CBM).
}
{
We fit GRB060614 light curves in Swift's BAT ($15-150$ keV) and XRT ($0.2-10$ keV) energy bands. Within the fireshell model, light curves are formed by two well defined and different components: the proper-GRB (P-GRB), emitted when the fireshell becomes transparent, and the extended afterglow, due to the interaction between the leftover accelerated baryonic and leptonic shell and the CBM.
}
{
We determine the two free parameters describing the GRB source within the fireshell model: the total $e^\pm$ plasma energy ($E_{tot}^{e^\pm} = 2.94\times 10^{51}$erg) and baryon loading ($B = 2.8\times 10^{-3}$). A small average CBM density $\sim 10^{-3}$ particles/cm$^3$ is inferred, typical of galactic halos. The first spikelike emission is identified with the P-GRB and the following prolonged emission with the extended afterglow peak. We obtain very good agreement in the BAT ($15-150$ keV) energy band, in what is traditionally called ``prompt emission'', and in the XRT ($0.2-10$ keV) one.
}
{
The \textit{anomalous} GRB060614 finds a natural interpretation within our canonical GRB scenario: it is a ``disguised'' short GRB. The total time-integrated extended afterglow luminosity is greater than the P-GRB one, but its peak luminosity is smaller since it is deflated by the peculiarly low average CBM density of galactic halos. This result points to an old binary system, likely formed by a white dwarf and a neutron star, as the progenitor of GRB060614 and well justifies the absence of an associated SN Ib/c. Particularly important for further studies of the final merging process are the temporal structures in the P-GRB down to $0.1$ s.
}


\maketitle 

\section{Introduction}

GRB060614 \citep{2006Natur.444.1044G,2007A&A...470..105M} has drawn the general attention of the gamma-ray burst's (GRB) scientific community because it is the first clear example of a nearby ($z=0.125$), long GRB not associated with a bright Ib/c supernova (SN) \citep{2006Natur.444.1050D,2006Natur.444.1053G}. It has been estimated that, if present, the SN-component should be about $200$ times fainter than the archetypal SN 1998bw associated with GRB980425; moreover, it would also be fainter (at least $30$ times) than any stripped-envelope SN ever observed \citep{2006AJ....131.2233R}.

Within the standard scenario, long duration GRBs ($T_{90} > 2$ s) are thought to be produced by SN events during the collapse of massive stars in star forming regions \citep[``collapsar'', see][]{1993ApJ...405..273W}. The observations of broad-lined and bright type Ib/c SNe associated with GRBs are often reported to favor this scenario \citep[][and references therein]{2006ARA&A..44..507W}. The \emph{ansatz} has been advanced that every long GRB should have a SN associated with it \citep{2007ApJ...655L..25Z}. Consequently, in all nearby long GRBs ($z \leq 1$), SN emission should be observed.

For these reasons the case of GRB060614 is unusual. Some obvious hypotheses have been proposed and ruled out: the chance superposition with a galaxy at low redshift \citep{2006Natur.444.1053G} and strong dust obscuration and extinction \citep{2006Natur.444.1047F}. Appeal has been made to the possible occurrence of an unusually low luminosity stripped-envelope core-collapse SN \citep{2006Natur.444.1050D}.

The second novelty of GRB060614 is that it challenges the traditional separation between Long Soft GRBs and Short Hard GRBs. Traditionally \citep{1992grbo.book..161K,1992AIPC..265..304D}, the ``short'' GRBs have $T_{90} < 2$ s, present an harder spectrum and negligible spectral lag, and are assumed to originate from the merging of two compact objects, i.e. two neutron stars or a neutron star and a black hole \citep[see e.g.][and references therein]{1984SvAL...10..177B,1986ApJ...308L..43P,1986ApJ...308L..47G,1989Natur.340..126E,2005RvMP...76.1143P,2006RPPh...69.2259M}. GRB060614 lasts about one hundred seconds \citep[$T_{90}=(102 \pm 5)$ s;][]{2006Natur.444.1044G}, it fulfills the $E_{p}^{rest}-E_{iso}$ correlation \citep{2007A&A...463..913A}, and therefore traditionally it should be classified as a ``long'' GRB. However, its morphology is different from typical long GRBs, being similar to the one of GRB050724, traditionally classified as a short GRB \citep{2007ApJ...655L..25Z,2005Natur.437..822P}. Its optical afterglow luminosity is intermediate between the traditional long and short ones \citep{2008arXiv0804.1959K}. Its host galaxy has a moderate specific star formation rate \citep[$R_{Host}\approx2M_{s}y^{-1}(L^{*})^{-1}$, $M_{vHost}\approx-15.5$;][]{2006Natur.444.1047F,2006Natur.444.1050D}. The spectral lag in its light curves is very small or absent \citep{2006Natur.444.1044G}. All these features are typical of short GRBs.

A third peculiarity of GRB060614 is that its $15$--$150$ keV light curve presents a short, hard and multi-peaked episode (about $5$ s). The episode is followed by a softer, prolonged emission that manifests a strong hard to soft evolution in the first $400$ s of data \citep{2007A&A...470..105M}. The total fluence in the $15$--$150$ keV energy band is $F=(2.17\pm0.04)\times10^{-5}$ erg/cm$^2$, the 20\% emitted during the initial spikelike emission, where the peak luminosity reaches the value of $300$ keV before decreasing to $8$ keV during the BAT-XRT overlap time (about $80$ s).

These apparent contradictions find a natural explanation in the framework of the ``fireshell'' model\footnote{We indicate here and in the following by the ``fireshell'' model the one we have consistently developed encompassing the three basic paradigms presented in \citet{2001ApJ...555L.107R,2001ApJ...555L.113R,2001ApJ...555L.117R} as well as all the technical developments in the emission process, in the equations of motion and in the relativistic treatment of the extended afterglow as summarized in \citet{2007AIPC..910...55R}.}. Within the fireshell model, the occurrence of a GRB-SN is not a necessity. The origin of all GRBs is traced back to the formation of a black hole, either occurring in a single process of gravitational collapse, or in a binary system composed of a neutron star and a companion star evolved out of the main sequence, or in a merging binary system composed of neutron stars and/or white dwarfs in all possible combinations. The occurrence of a GRB-SN is indeed only one of the possibilities, linked, for example, to the process of ``induced gravitational collapse'' \citep{2001ApJ...555L.117R,Mosca_Orale,2007A&A...471L..29D}.

Within the ``fireshell'' model, the difference between the ``short'' and ``long'' GRBs does not rely uniquely on the time scale of the event, as in the traditional classification. It is theoretically explained by the fireshell baryon loading affecting the structure of the ``canonical'' GRB light curve. The ``canonical'' GRB light curve is indeed composed of a proper-GRB (P-GRB, often labeled as ``precursor''), emitted at the fireshell transparency, and an extended afterglow. The peak of such an extended afterglow is traditionally included in the prompt emission. The relative energetics of two such components and the temporal separation of the corresponding peaks are ruled by the fireshell baryon loading. In the limit of vanishing baryon loading, all the GRB energy is emitted in the P-GRB: these are the ``genuine'' short GRBs \citep{2001ApJ...555L.113R,2008AIPC.1065..219R,2007A&A...474L..13B,2008AIPC..966...12B,2008AIPC.1000..305B}.

Within the ``fireshell'' model, in addition to the determination of the baryon loading, it is possible to infer a detailed description of the circumburst medium (CBM), its average density and its porosity and filamentary structure, all the way from the black hole horizon to a distance $r \lesssim 10^{17}$ cm. This corresponds to the prompt emission. This description is lacking in the traditional model based on the synchrotron emission. The attempt to use the internal shock model for the prompt emission \citep[see e.g.][and references therein]{1994ApJ...430L..93R,2005RvMP...76.1143P,2006RPPh...69.2259M} only applies to regions where $r > 10^{17}$ cm \citep{2008MNRAS.384...33K}.

The aim of this paper is to show how the ``fireshell'' model can explain all the abovementioned GRB060614 peculiarities and solve the apparent contradictions. In doing so, we also infer constraints on the astrophysical nature of the GRB060614 progenitors. In turn, these conclusions lead to a new scenario for all GRBs. We can confirm a classification of GRBs as ``genuine'' short, ``fake'' or ``disguised'' short, and all the remaining ``canonical'' GRBs. The connection between this new classification and the nature of GRB progenitors is quite different from the traditional one in the current literature.

In Sect. \ref{model} we recall the main features of the fireshell model relevant to the present analysis. In Sect. \ref{progenitors} we present the astrophysical scenario inferred by the ``fireshell'' model for the nature of GRB progenitors. In Sect. \ref{fit} we present and discuss the fit of the gapless GRB060614 light curves provided by the Swift satellite \citep{2004ApJ...611.1005G} in the BAT ($15-150$ keV) energy band, in what is traditionally called ``prompt emission'', and in the XRT ($0.2-10$ keV) one, in the entire afterglow. In Sect. \ref{open_issues} we discuss some still open issues in the theoretical analysis. In Sect. \ref{conclusions} our general conclusions are presented.

\section{Brief reminder of the \textit{fireshell} model}\label{model}

The black hole uniqueness theorem \citep[see e.g.][]{1971PhT....24a..30R} is at the basis of the fact that it is possible to explain the different GRB features with a single theoretical model, over a range of energies spanning over $6$ orders of magnitude. The fundamental point is that, independently of the fact that the progenitor of the gravitational collapse is represented by merging binaries composed of neutron stars and white dwarfs in all possible combinations, or by a single process of gravitational collapse, or by the process of ``induced'' gravitational collapse, the formed black hole is totally independent of the initial conditions and reaches the standard configuration of a Kerr-Newman black hole. It is well known that pair creation by the vacuum polarization process can occur in a Kerr-Newman black hole \citep{1975PhRvL..35..463D,PhysRep}.

We consequently assume, within the fireshell model, that all GRBs originate from an optically thick $e^\pm$ plasma with total energy $E_{tot}^{e^\pm}$ in the range $10^{49}$--$10^{54}$ ergs and a temperature $T$ in the range $1$--$4$ MeV \citep{1998A&A...338L..87P}. Such an $e^\pm$ plasma has been widely adopted in the current literature \citep[see e.g.][and references therein]{2005RvMP...76.1143P,2006RPPh...69.2259M}. After an early expansion, the $e^\pm$-photon plasma reaches thermal equilibrium with the engulfed baryonic matter $M_B$ described by the dimensionless parameter $B=M_{B}c^{2}/E_{tot}^{e^\pm}$, that must be $B < 10^{-2}$ \citep{1999A&A...350..334R,2000A&A...359..855R}. As the optically thick fireshell composed of $e^\pm$-photon-baryon plasma self-accelerates to ultrarelativistic velocities, it finally reaches the transparency condition. A flash of radiation is then emitted. This is the P-GRB \citep{2001ApJ...555L.113R}. Different current theoretical treatments of these early expansion phases of GRBs are compared and contrasted in \citet{brvx06} and \citet{2008AIPC.1065..219R}. The amount of energy radiated in the P-GRB is only a fraction of the initial energy $E_{tot}^{e^\pm}$. The remaining energy is stored in the kinetic energy of the optically thin baryonic and leptonic matter fireshell that, by inelastic collisions with the CBM, gives rise to a multi-wavelength emission. This is the extended afterglow. It presents three different regimes: a rising part, a peak and a decaying tail. What is usually called ``prompt emission'' in the current literature mixes the P-GRB with the raising part and the peak of the extended afterglow. Such an unjustified mixing of these components, originating from different physical processes, leads to difficulties in the current models of GRBs, and can as well be responsible for some of the intrinsic scatter observed in the Amati relation \citep{2006MNRAS.372..233A,2008A&A...487L..37G}.

\begin{figure}
\includegraphics[width=\hsize]{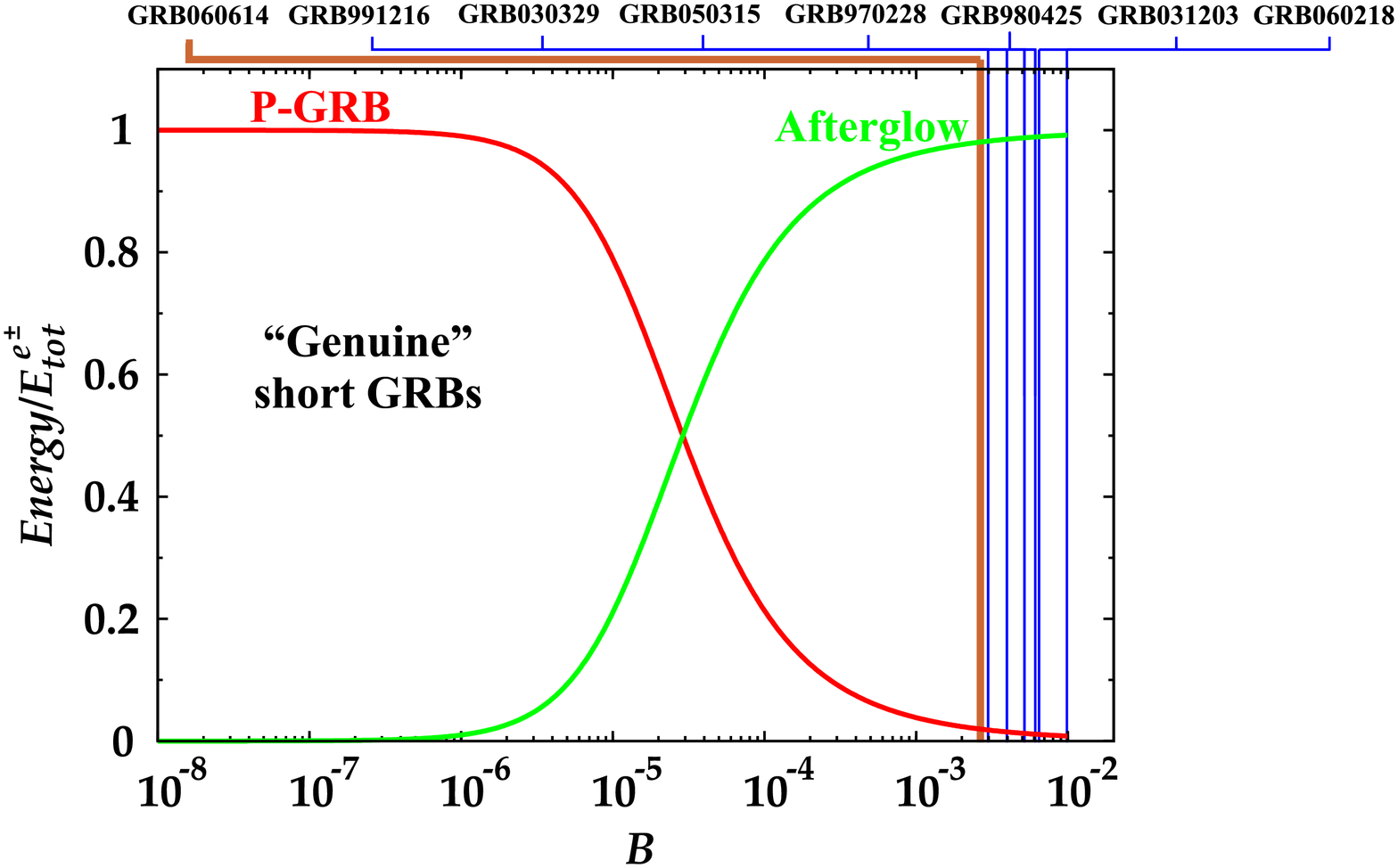}
\caption{Here the energies emitted in the P-GRB (red line) and in the extended afterglow (green line), in units of the total energy of the plasma, are plotted as functions of the $B$ parameter. When $B \lesssim 10^{-5}$, the P-GRB becomes predominant over the extended afterglow, giving rise to a ``genuine'' short GRB. In the figure are also marked in blue the values of the $B$ parameters corresponding to some GRBs we analyzed, all belonging to the class of long GRBs, together with the GRB060614 one (thick brown line).}
\label{f2}
\end{figure}

At the transparency point, the value of the $B$ parameter rules the ratio between the energetics of the P-GRB and the kinetic energy of the baryonic and leptonic matter giving rise to the extended afterglow. It rules as well the time separation between the corresponding peaks \citep{2001ApJ...555L.113R,2008AIPC.1065..219R}. Within our classification a canonical GRB for baryon loading $B \lesssim 10^{-5}$ has the P-GRB component energetically dominant over the extended afterglow (see Fig. \ref{f2}). In the limit $B \rightarrow 0$ it gives rise to a ``genuine'' short GRB. Otherwise, when $10^{-4} \lesssim B \leq 10^{-2}$, the kinetic energy of the baryonic and leptonic matter, and consequently the emission of the extended afterglow, is dominant with respect to the P-GRB \citep{2001ApJ...555L.113R,2008AIPC.1065..219R,2007A&A...474L..13B,2008AIPC.1065..223B}.

The extended afterglow luminosity in the different energy bands is governed by two quantities associated with the environment. Within the fireshell model, these are the effective CBM density profile, $n_{cbm}$, and the ratio between the effective emitting area $A_{eff}$ and the total area $A_{tot}$ of the expanding baryonic and leptonic shell, ${\cal R}= A_{eff}/A_{tot}$. This last parameter takes into account the CBM filamentary structure \citep{2004IJMPD..13..843R,2005IJMPD..14...97R} and the possible occurrence of a fragmentation in the shell \citep{2007A&A...471L..29D}. In our hypothesis, the emission from the baryonic and leptonic matter shell is spherically symmetric. This allows us to assume, in a first approximation, a modeling of thin spherical shells for the CBM distribution and consequently to consider just its radial dependence \citep{2002ApJ...581L..19R}. The emission process is postulated to be thermal in the co-moving frame of the shell \citep{2004IJMPD..13..843R}. The observed GRB non-thermal spectral shape is due to the convolution of an infinite number of thermal spectra with different temperatures and different Lorentz and Doppler factors. Such a convolution is to be performed over the surfaces of constant arrival time of the photons at the detector \citep[EQuiTemporal Surfaces, EQTSs;][]{2005ApJ...620L..23B} encompassing the whole observation time \citep{2005ApJ...634L..29B}.

\begin{figure}
\includegraphics[width=\hsize,clip]{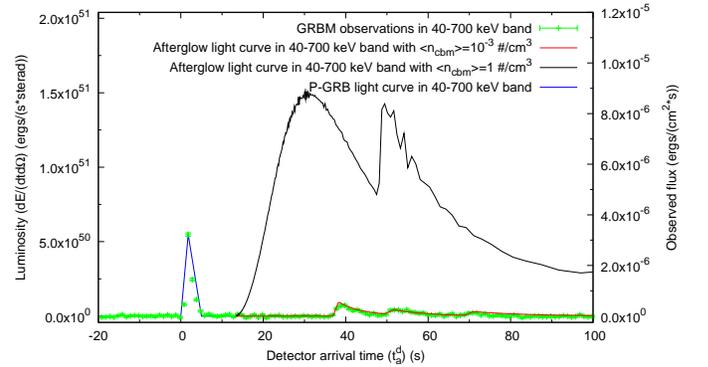}
\caption{The theoretical fit of the \emph{Beppo}SAX GRBM observations of GRB970228 in the $40$--$700$ keV energy band. The red line corresponds to an average CBM density $\sim 10^{-3}$ particles/cm$^3$. The black line is the extended afterglow light curve obtained rescaling the CBM density to $\langle n_{cbm} \rangle = 1$ particle/cm$^3$ keeping constant its shape and the values of $E_{e^\pm}^{tot}$ and $B$. The blue line is the P-GRB. Details in \citet{2007A&A...474L..13B}.}
\label{picco_n=1}
\end{figure}

In the present paper, we extend the theoretical understanding, within the fireshell model, of a new class of sources, pioneered by \citet{2006ApJ...643..266N}, of which the present GRB060614 is clearly a member. This class is characterized by an occasional softer extended emission after an initial spikelike emission. The softer extended emission has a peak luminosity smaller than the one of the initial spikelike emission. This has misled the understanding of the correct role of the extended afterglow, as summarized in the introduction. As shown in the prototypical case of GRB970228 \citep{2007A&A...474L..13B}, the initial spikelike emission can be identified with the P-GRB and the softer extended emission with the peak of the extended afterglow. The fact that the time-integrated extended afterglow luminosity is much larger than the P-GRB one is crucial, and this fact unquestionably identifies GRB970228 as a canonical GRB with $B > 10^{-4}$. The consistent application of the fireshell model allowed us to compute the CBM porosity, filamentary structure and average density which, in that specific case, was $n_{cbm} \sim 10^{-3}$ particles/cm$^3$ \citep{2007A&A...474L..13B}. This explained the peculiarity of the low extended afterglow peak luminosity and of its much longer time evolution. These features are not intrinsic to the progenitor nor to the black hole, but they uniquely depend on the peculiarly low value of the CBM density, typical of galactic halos. If one takes the same total energy, baryon loading and CBM distribution as in GRB970228, and rescales the CBM density profile by a constant numerical factor in order to raise its average value from $10^{-3}$ to $1$ particles/cm$^3$, one obtains a GRB with a much larger extended afterglow peak luminosity and a much reduced time scale. Such a GRB would appear a perfect traditional ``long'' GRB following the current literature \citep[see Fig. \ref{picco_n=1} and][]{2007A&A...474L..13B}. This has led us to expand the traditional classification of GRBs to three classes: ``genuine'' short GRBs, ``fake'' or ``disguised'' short GRBs, and all the remaining ``canonical'' ones \citep{RuCef}.

\section{The ``fireshell'' model and GRB progenitors}\label{progenitors}

We have recalled in the introduction that ``long'' GRBs are traditionally related in the current literature to the idea of a single progenitor, identified as a ``collapsar'' \citep{1993ApJ...405..273W}. Similarly, short GRBs are assumed to originate from binary mergers formed by white dwarfs, neutron stars, and black holes in all possible combinations. It also has been suggested that short and long GRBs originate from different galaxy types. In particular, short GRBs are proposed to be associated with galaxies with low specific star forming rate \citep[see e.g.][]{2009ApJ...690..231B}. Some evidence against such a scenario have been advanced, due to the small sample size and the different estimates of the star forming rates \citep[see e.g.][]{2009ApJ...691..182S}. However, the understanding of GRB structure and of its relation to the CBM distribution, within the fireshell model, leads to a more complex and interesting perspective than the one in the current literature.

The first general conclusion of the ``fireshell'' model \citep{2001ApJ...555L.113R} is that, while the time scale of ``short'' GRBs is indeed intrinsic to the source, this does not happen for the ``long'' GRBs: their time scale is clearly only a function of the instrumental noise threshold. This has been dramatically confirmed by the observations of the Swift satellite \citep{Venezia_Orale}. Among the traditional classification of ``long'' GRBs we distinguish two different sub-classes of events, neither of which originates from collapsars.

The first sub-class contains ``long'' GRBs that are particularly weak ($E_{iso} \sim 10^{50}$ erg) and associated with SN Ib/c. In fact, it has been often proposed that such GRBs, only observed at smaller redshift $0.0085 < z < 0.168$, form a different class, less luminous and possibly much more numerous than the high luminosity GRBs at higher redshift \citep{2006Natur.442.1011P,2004Natur.430..648S,2007ApJ...658L...5M,2006AIPC..836..367D}. Therefore in the current literature they have been proposed to originate from a separate class of progenitors \citep{2007ApJ...662.1111L,2006ApJ...645L.113C}. Within our ``fireshell'' model, they originate in a binary system formed by a neutron star, close to its critical mass, and a companion star, evolved out of the main sequence. They produce GRBs associated with SNe Ib/c, via the ``induced gravitational collapse'' process \citep{2001ApJ...555L.117R}. The low luminosity of these sources is explained by the formation of a black hole with the smallest possible mass: the one formed by the collapse of a just overcritical neutron star \citep{Mosca_Orale,2007A&A...471L..29D}.

A second sub-class of ``long'' GRBs originates from merging binary systems, formed either by two neutron stars or a neutron star and a white dwarf. A prototypical example of such systems is GRB970228. The binary nature of the source is inferred by its migration from its birth location in a star forming region to a low density region within the galactic halo, where the final merging occurs \citep{2007A&A...474L..13B}. The location of such a merging event in the galactic halo is indeed confirmed by optical observations of the GRB970228 afterglow  \citep{1997Natur.387R.476S,1997Natur.386..686V}. The crucial point is that, as recalled above, GRB970228 is a ``canonical'' GRB with $B > 10^{-4}$ ``disguised'' as a short GRB. We are going to see in the following that GRB060614 also comes from such a progenitor class.

If the binary merging would occur in a region close to its birth place, with an average density of $1$ particle/cm$^3$, the GRB would appear as a traditional high-luminosity ``long'' GRB, of the kind currently observed at higher redshifts (see Fig. \ref{picco_n=1}), similar to, e.g., GRB050315 \citep{2006ApJ...645L.109R}.

Within our approach, therefore, there is the distinct possibility that all GRB progenitors are formed by binary systems, composed of neutron stars, white dwarfs, or stars evolved out of the main sequence, in different combinations.

The case of the ``genuine'' short GRBs is currently being examined within the ``fireshell'' model.

\section{The fit of the observed luminosity}\label{fit}

\begin{figure}
\includegraphics[width=\hsize]{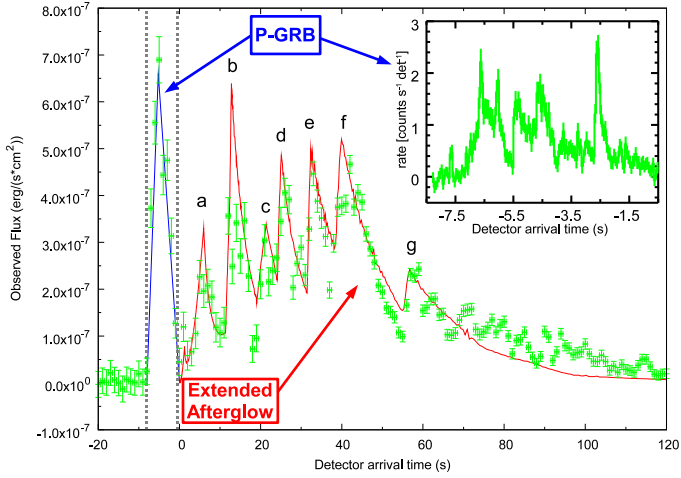}
\caption{The BAT $15$--$150$ keV light curve (green points) at $1$ s time resolution compared with the corresponding theoretical extended afterglow light curve we compute (red line). The onset of the extended afterglow is at the end of the P-GRB (qualitatively sketched in blue lines and delimited by dashed gray vertical lines). Therefore the zero of the temporal axis is shifted by $5.5$ s with respect to the BAT trigger time. The peaks of the extended afterglow light curves are labeled to match them with the corresponding CBM density peak in Fig. \ref{f4a}. In the upper right corner there is an enlargement of the P-GRB at $50$ms time resolution \citep[reproduced from][]{2007A&A...470..105M}, showing its structure.}
\label{f2a}
\end{figure}

\begin{figure}
\includegraphics[width=\hsize]{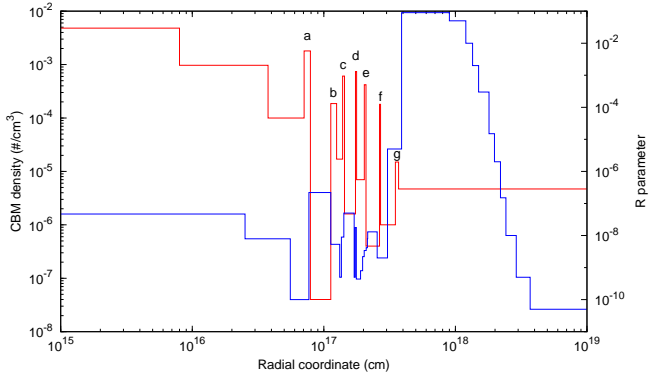}
\caption{The effective CBM density (red line) and the ${\cal R}$ parameter (blue line) versus the radial coordinate of the shell. The CBM density peaks are labeled to match them with the corresponding extended afterglow light curve peaks in Fig. \ref{f2a}. They correspond to filaments of characteristic size $\Delta r \sim 10^{15}$ cm and density contrast $\Delta n_{cbm}/\langle n_{cbm} \rangle \sim 20$ particles/cm$^3$.}
\label{f4a}
\end{figure}

\begin{figure}
\includegraphics[width=\hsize]{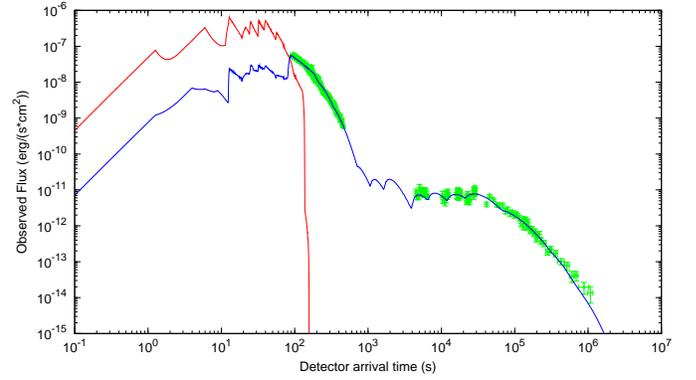}
\caption{The XRT $0.2$--$10$ keV light curve (green points) compared with the corresponding theoretical extended afterglow light curve we compute (blue line). In this case also we have a good correspondence between data and theoretical results. For completeness, the red line shows the theoretical extended afterglow light curve in the $15$--$150$ keV energy range presented in Fig. \ref{f2a}.}
\label{f4}
\end{figure}

In this scenario, GRB060614 is naturally interpreted as a ``disguised'' short GRB. We have performed the analysis of the observed light curves in the $15$--$150$ keV energy band, corresponding to the $\gamma$-ray emission observed by the BAT instrument on the Swift satellite, and in the $0.2$--$10$ keV energy band, corresponding to the X-ray component from the XRT instrument on Swift satellite. We do not address in this paper the issue of the optical emission, which represent less than 10\% of the total energy of the GRB. From this fit (see Figs. \ref{f2a}, \ref{f4}) we have derived the total initial energy $E_{tot}^{e^\pm}$, the value of $B$ as well as the effective CBM distribution (see Fig. \ref{f4a}). We find $E_{tot}^{e^\pm}=2.94\times10^{51}$ erg, that accounts for the bolometric emission of both the P-GRB and the extended afterglow. Such a value is compatible with the observed $E_{iso} \simeq 2.5\times10^{51}$ erg \citep{2006Natur.444.1044G}. The value of $B$ is $B=2.8\times10^{-3}$, which corresponds to the lowest one of all the GRBs we have examined (see Fig. \ref{f2}). It corresponds to a canonical GRB with a very clear extended afterglow predominance over the P-GRB. From the model, having determined $E_{tot}^{e^\pm}$ and $B$, we can compute the theoretical expected P-GRB energetics $E_{P-GRB}$ \citep{2001ApJ...555L.113R}. We obtain $E_{P-GRB} \simeq 1.15 \times 10^{50}$ erg, that is in good agreement with the observed $E_{iso,1p} \simeq 1.18\times10^{50}$ erg \citep{2006Natur.444.1044G}. The Lorentz Gamma Factor at the transparency is $\gamma_\circ=346$, one of the highest of all the GRBs we have examined.

In Fig. \ref{f2a} we plot the comparison between the BAT observational data of the GRB0606014 prompt emission in the $15$--$150$ keV energy range and the P-GRB and extended afterglow light curves computed within our model. The temporal variability of the extended afterglow peak emission is due to the inhomogeneities in the effective CBM density (see Figs. \ref{f2a}, \ref{f4a}). Toward the end of the BAT light curve, the good agreement between the observations and the fit is affected by the Lorentz gamma factor decrease and the corresponding increase of the maximum viewing angle. The source visible area becomes larger than the typical size of the filaments. This invalidates the radial approximation we use for the CBM description. To overcome this problem it is necessary to introduce a more detailed three-dimensional CBM description, in order to avoid an over-estimated area of emission and, correspondingly, to describe the sharpness of some observed light curves. We are still working on this issue \citep{2002ApJ...581L..19R,C07,Venezia_Flares,G07}.

We turn now to the crucial determination of the CBM density, which is derived from the fit. At the transparency point it was $n_{cbm} = 4.8 \times 10^{-3}$ particles/cm$^3$ (see Fig. \ref{f4a}). This density is compatible with the typical values of the galactic halos. During the peak of the extended afterglow emission the effective average CBM density decreases reaching $\left\langle n_{cbm}\right\rangle = 2.25 \times 10^{-5}$ particles/cm$^3$, possibly due to an occurring fragmentation of the shell \citep{2007A&A...471L..29D} or due to a fractal structure in the CBM. The ${\cal R}$ value on average was $\left\langle {\cal R}\right\rangle = 1.72 \times 10^{-8}$. Note the striking analogy of the numerical value and the overall radial dependence of the CBM density in the present case of GRB060614 when compared and contrasted with the ones of GRB970228 \citep{2007A&A...474L..13B}.

Concerning the $0.2$--$10$ keV light curve of the decaying phase of the afterglow observed with the XRT instrument, we have also reproduced very satisfactorily both the hard decrease in the slope and the plateau of the light curve, keeping constant the effective CBM density and changing only ${\cal R}$. The result of this analysis is reported in Fig. \ref{f4}. We assume in this phase $n_{cbm} = 4.70 \times 10^{-6}$ particles/cm$^{-3}$. The average value of the ${\cal R}$ parameter is $\left\langle {\cal R}\right\rangle=1.27\times10^{-2}$. The drastic enhancement in the ${\cal R}$ parameter with respect to the values at the peak of the extended afterglow is consistent with similar features encountered in other sources we have studied: GRB060218 presents an enhancement of five orders of magnitude \citep{2007A&A...471L..29D}, in GRB060710 the enhancement is of about four orders of magnitude (see Izzo et al., in preparation) while in GRB050315 there is a three orders of magnitude enhancement \citep{2006ApJ...645L.109R}. In these last two cases, we find the enhancement of ${\cal R}$ between $r$=$2\times10^{17}$ cm and $r$=$3\times10^{17}$ cm, just as for GRB060614, for which we have the enhancement at $r$=$3.5\times10^{17}$ cm. The time of the bump approximately corresponds to the appearance of the optical emission observed in GRB060614 and, more in general, to the onset of the second component of the \citet{2007ApJ...662.1093W} scheme for GRBs.
 
\section{Open issues in current theoretical models}\label{open_issues} 

The ``fireshell'' model addresses mainly the $\gamma$ and X-ray emission, which are energetically the most relevant part of the GRB phenomenon. The model allows a detailed identification of the fundamental three parameters of the GRB source: the total energy, the baryon loading, as well as the CBM density, filamentary structure and porosity. Knowledge of these phenomena characterizes the region surrounding the black hole up to a distance which in this source reaches $\sim 10^{17}$--$10^{18}$ cm. When applied, however, to larger distances, which correspond to the latest phases of the X-ray afterglow, since the beginning of the ``plateau'' phase, the model reveals a different regime which has not yet been fully interpreted in its astrophysical implications. To fit the light curve in the soft X-ray regime for $r > 4 \times 10^{17}$ cm, we require an enhancement of about six orders of magnitude in the ${\cal R}$ factor (see Fig. \ref{f4a}). This would correspond to a more diffuse CBM structure, with a smaller porosity, interacting with the fireshell. This implies a different main physical process during the latest X-ray afterglow phases. For the optical, IR and radio emission, the fireshell model leads to a much lower flux than the observed one, especially for $r \sim 10^{17}$--$10^{18}$ cm. Although the optical, IR, and radio luminosities have a minor energetic role, they may lead to the identification of crucial parameters and new phenomena occurring in the source, and they deserve careful attention.

In these late phases for $r \geq 10^{17}$ cm the treatment based on synchrotron emission, pioneered by \citet{1997ApJ...476..232M} even before the discovery of the afterglow \citep{1997Natur.387..783C}, is currently applied. Such a model has been further developed \citep[see][and references therein]{1998ApJ...497L..17S,2005RvMP...76.1143P,2006RPPh...69.2259M}. Also in this case, however, some difficulties remain since it is necessary to invoke the presence of an unidentified energy injection mechanism \citep{2006ApJ...642..354Z}. Such a model appears to be quite successful in explaining the late phases of the X-ray emission of GRB060614, as well as the corresponding optical emission, in terms of different power-law indexes for the different parts of the afterglow light curves \citep{2007A&A...470..105M,2008arXiv0812.0979X}. However, also in this case an unidentified energy injection mechanism between $\sim 0.01$ days and $\sim 0.26$ days appears to be necessary \citep{2008arXiv0812.0979X}. 

The attempt to describe the prompt emission via the synchrotron process by the internal shock scenario \citep[see e.g.][and references therein]{1994ApJ...430L..93R,2005RvMP...76.1143P,2006RPPh...69.2259M} also encounters difficulties: \citet{2008MNRAS.384...33K} have shown that the traditional synchrotron model can be applied to the prompt emission only if it occurs at $r > 10^{17}$ cm. A proposed solution to this problem, via the inverse Compton process, suffers from an ``energy crisis'' \citep[see e.g.][]{2008arXiv0807.3954P}. 

Interestingly, the declared region of validity of the traditional synchrotron model ($r > 10^{17}$ cm) is complementary to the one successfully described by our model ($r < 10^{17}$--$10^{18}$ cm). Astrophysically, \citet{2008arXiv0812.0979X} have reached, within the framework of the traditional synchrotron model, two conclusions which are consistent with the results of our analysis. First, they also infer from their numerical fit a very low density environment, namely $n_{cbm} \sim 0.04$ particles/cm$^3$. Second, they also mention the possibility that the progenitor of GRB0606014 is a merging binary system formed by two compact objects. 

\section{Conclusions}\label{conclusions}

As recalled in the introduction, GRB060614 presents three major novelties, which challenge the most widespread theoretical models and which are strongly debated in the current literature. The first one is that it challenges the traditional separation between long soft GRBs and short hard GRBs \citep{2006Natur.444.1044G}. The second one is that it presents a short, hard and multi-peaked episode, followed by a softer, prolonged emission with a strong hard to soft evolution \citep{2006Natur.444.1044G,2007A&A...470..105M}. The third one is that it is the first clear example of a nearby, long GRB not associated with a bright SN Ib/c \citep{2006Natur.444.1050D,2006Natur.444.1053G}. All these three issues are naturally explained within our ``fireshell'' model, which allows a detailed analysis of the temporal behavior of the signal originating up to a distance $r \sim 10^{17}$--$10^{18}$ cm from the black hole, and relates, with all the relativistic transformation, the arrival time to the CBM structure and the relativistic parameters of the fireshell.

One of the major outcomes of the Swift observation of, e.g., GRB050315 \citep{2006ApJ...638..920V,2006ApJ...645L.109R} has been the confirmation that long GRB duration is not intrinsic to the source but it is merely a function of the instrumental noise threshold \citep{Venezia_Orale}. GRB060614 represents an additional fundamental progress in clarifying the role of the CBM density in determining the GRB morphology. It confirms the results presented in GRB970228 \citep{2007A&A...474L..13B}, that is the prototype of the new class of ``fake'' short GRBs, or, better, of canonical GRBs ``disguised'' as short ones. They correspond to canonical GRBs with an extended afterglow emission energetically predominant relative to the P-GRB one and a baryon loading $B > 10^{-4}$. The sharp spiky emission corresponds to the P-GRB. As recalled in the introduction and in Sect. \ref{model}, a comparison of the luminosities of the P-GRB and of the extended afterglow is indeed misleading: it follows from the low average CBM density inferred from the fit of the fireshell model, which leads to $n_{cbm} \sim 10^{-3}$ particles/cm$^3$. Therefore such a feature is neither intrinsic to the progenitor nor to the black hole, but it is only indicative of the CBM density at the location where the final merging occurs. GRB060614 is a canonical GRB and it is what would be traditionally called a ``long'' GRB if it had not exploded in an especially low CBM density environment. GRB060614 must necessarily fulfill, and indeed it does, the Amati relation. This happens even taking into account the entire prompt emission mixing together the P-GRB and the extended afterglow \citep{2007A&A...463..913A}, due to the energetic predominance of the extended afterglow discussed above \citep[see also][]{2008A&A...487L..37G}. These results justify the occurrence of the abovementioned first two novelties.

The low value of the CBM density is compatible with a galactic halo environment. This result points to an old binary system as the progenitor of GRB060614 and it justifies the abovementioned third novelty: the absence of an associated SN Ib/c \citep[see also][]{2007AIPC..906...69D}. Such a binary system departed from its original location in a star-forming region and spiraled out in a low density region of the galactic halo \citep[see e.g.][]{KMG11}. The energetics of this GRB is about two orders of magnitude lower than the one of GRB970228 \citep{2007A&A...474L..13B}. A natural possible explanation is that instead of a neutron star - neutron star merging binary system we are in the presence of a white dwarf - neutron star binary. We therefore agree, for different reasons, with the identification proposed by \citet{2007AIPC..906...69D} for the GRB060614 progenitor. In principle, the nature of the white dwarf, with a typical radius of the order of $10^3$ km, as opposed to the one of the neutron star, typically of the order of $10$ km, may manifest itself in characteristic signatures in the structure of the P-GRB (see Fig. \ref{f2a}).

It is interesting that these results lead also to three major new possibilities:
\begin{itemize}
\item The majority of GRBs declared as shorts \citep[see e.g.][]{2005Natur.437..822P} are likely ``disguised'' short GRBs, in which the extended afterglow is below the instrumental threshold.
\item The observations of GRB060614 offer the opportunity, for the first time, to analyze in detail the structure of a P-GRB lasting $5$ s. This feature is directly linked to the physics of the gravitational collapse that generated the GRB. Recently, there has been a crucial theoretical physics result showing that the characteristic time constant for the thermalization of an $e^\pm$ plasma is of the order of $10^{-13}$ s \citep{2007PhRvL..99l5003A}. Such a time scale still applies for an $e^\pm$ plasma with a baryon loading of the order of the one observed in GRBs \citep{PRD}. The shortness of such a time scale, as well as the knowledge of the dynamical equations of the optically thick phase preceding the P-GRB emission \citep{brvx06}, implies that the structure of the P-GRB is a faithful representation of the gravitational collapse process leading to the formation of the black hole \citep{2005IJMPD..14..131R}. In this respect, it is indeed crucial that the Swift data on the P-GRB observed in GRB060614 (see Fig. \ref{f2a}) appear to be highly structured all the way to a time scale of $0.1$ s. This opens a new field of research: the study of the P-GRB structure in relation to the process of gravitational collapse leading to the GRB.
\item If indeed the binary nature of the progenitor system and the peculiarly low CBM density $n_{cbm} \sim 10^{-3}$ particles/cm$^3$ will be confirmed for all ``fake'' or ``disguised'' GRBs, then it is very likely that the traditionally ``long'' high luminosity GRBs at higher redshift also originates in the merging of binary systems formed by neutron stars and/or white dwarfs occurring close to their birth location in star-forming regions with $n_{cbm} \sim 1$ particle/cm$^3$ (see Fig. \ref{picco_n=1}).
\end{itemize}

\acknowledgements

We thank the Italian Swift Team (supported by ASI grant I/011/07/0 and partly by the MIUR grant 2005025417) for the reduced Swift data, as well as an anonymous referee for her/his stimulating discussions, suggestions and advice.

\end{document}